\documentclass[a4paper,11pt]{article}

\usepackage{pos}

\title{Off-shell effects in $t\bar{t}+\gamma/Z$ production at the LHC}
\ShortTitle{Off-shell effects in $t\bar{t}+\gamma/Z$ production at the LHC}

\author*[\dag]{Giuseppe Bevilacqua} 
\notes{\note{Work supported by grant K 125105 of the National Research, Development and Innovation Office in Hungary.}}

\affiliation{MTA-DE Particle Physics Research Group \\ H-4002 Debrecen, PO Box 400, Hungary}

\emailAdd{giuseppe.bevilacqua@science.unideb.hu}
\abstract{
We present state-of-the-art predictions for $t\bar{t}\gamma$ and $t\bar{t}Z(Z \to \nu_\ell\bar{\nu}_\ell)$ production in the fully leptonic channel at the LHC. The first process can provide key information on the strength and structure of the top-quark coupling to the photon. The invisible-decay channel $t\bar{t}Z(Z \to \nu_\ell\bar{\nu}_\ell)$ is an important SM background for Dark Matter searches in $t\bar{t}+E_{T}^{miss}$ final states.
Our results are accurate to NLO in QCD and include all resonant and non-resonant diagrams, interferences and off-shell effects for top quarks and $W/Z$ bosons. As such, they represent the most complete description of these processes from the viewpoint of a fixed-order calculation. We show selected results for the LHC Run II energy of 13 TeV, focusing on the impact of choosing well motivated scales and on the effect of using different approaches for the modeling of top quark decays.
}

\FullConference{
  The Eighth Annual Conference on Large Hadron Collider Physics-LHCP2020 \\
  25-30 May, 2020\\
  online}

\begin{document}

\maketitle

\section{Introduction}

Given the total luminosity collected at the LHC during the Run II, top quark physics has entered a new precision era.
While top quark pairs are produced abundantly via strong interactions, also rarer associated channels such as $t\bar{t}+V (V=\gamma,Z)$ can be studied in some detail \cite{Aaboud:2018hip,Aad:2020axn,Aaboud:2019njj,Sirunyan:2017uzs,CMS:2019too,ATLAS:2020cxf}. This opens a window for precision measurements of top-quark gauge couplings, which are at the center of the interest of current analyses of anomalous couplings and EFT interpretations (see \textit{e.g.} \cite{Baur:2004uw,Bylund:2016phk,Schulze:2016qas,Brivio:2019ius,Bissmann:2019gfc}). Additionally,  $t\bar{t}Z$ with invisible $Z$-boson decays is an important background for Dark Matter (DM) searches in the $t\bar{t}+E_{T}^{miss}$ channel \cite{Haisch:2016gry}. On the theory side, continuous efforts have been devoted to improve the theoretical modeling of $t\bar{t}\gamma$ \cite{PengFei:2009ph,PengFei:2011qg,Duan:2016qlc,Maltoni:2015ena,Melnikov:2011ta,Kardos:2014zba} and $t\bar{t}Z$ \cite{Lazopoulos:2008de,Rontsch:2014cca,Garzelli:2011is,Frixione:2015zaa,Kulesza:2020nfh,Broggio:2019ewu} at the LHC. Typically, top quarks are treated as stable particles or decayed in the Narrow Width Approximation (NWA), depending whether the focus of the calculation is on inclusive production rates or fiducial cross sections. It is only quite recently that the first complete NLO QCD predictions for realistic final states, including off-shell and non-resonant effects, have started to appear for these processes \cite{Bevilacqua:2018woc,Bevilacqua:2018dny,Bevilacqua:2019quz,Bevilacqua:2019cvp}. In this proceedings we report a selection of results from our most recent analyses of $t\bar{t}Z(Z \to \nu_\ell\bar{\nu}_\ell)$  and $t\bar{t}\gamma$ with leptonic decays at 13 TeV, as presented in Ref.\cite{Bevilacqua:2019quz,Bevilacqua:2019cvp}.

\section{Predictions for $t\bar{t}Z(Z \to \nu_\ell\bar{\nu}_\ell)$ and $t\bar{t}\gamma$ with leptonic decays}

We study the processes $pp \to b \bar{b} e^+ \nu_e \mu^- \bar{\nu}_\mu \nu_\tau \bar{\nu}_\tau \, + X$ \cite{Bevilacqua:2019cvp} and $pp \to b \bar{b} e^+ \nu_e \mu^- \bar{\nu}_\mu \gamma \, + X$ \cite{Bevilacqua:2019quz} at the LHC energy of 13 TeV. For brevity, in the following we will refer to them as to "$t\bar{t}Z(Z \to \nu_\ell\bar{\nu}_\ell)$" and "$t\bar{t}\gamma$" respectively. All details of the calculational setup and kinematical cuts can be found in our published work.  Our results are accurate at NLO QCD accuracy and include all resonant and non-resonant Feynman diagrams, interferences and finite-width effects at fixed perturbative order. They have been obtained with the help of the package \textsc{Helac-Nlo} \cite{Bevilacqua:2011xh}, which consists of \textsc{Helac-1loop} \cite{vanHameren:2009dr} and \textsc{Helac-Dipoles} \cite{Czakon:2009ss,Bevilacqua:2013iha}. Events are stored in either Les Houches Event File format \cite{Alwall:2006yp} or ROOT Ntuples \cite{Antcheva:2009zz} that might be directly used for experimental studies. 
The ROOT Ntuples can be processed by a dedicated in-house C++ analysis framework, \textsc{Heplot}, to obtain  predictions for arbitrary infrared-safe observables, scale/PDF setups and kinematical cuts. The Ntuples and the analysis framework are both available upon request to the authors.

\begin{figure}[h!]
\begin{center}
\includegraphics[width=.42\textwidth, height=6.7cm]{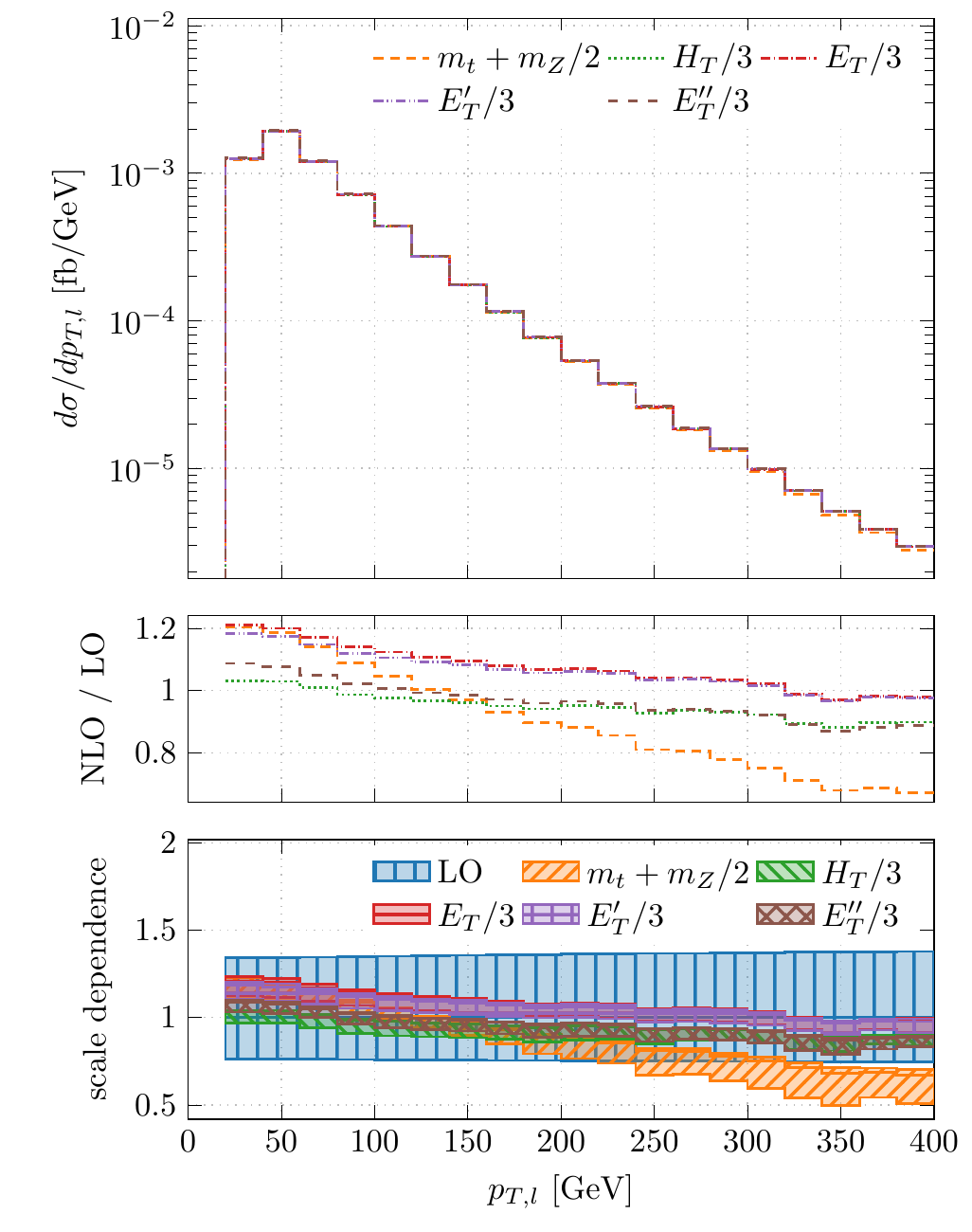} 
\hspace{0.3cm}
\includegraphics[width=.46\textwidth, height=6.7cm]{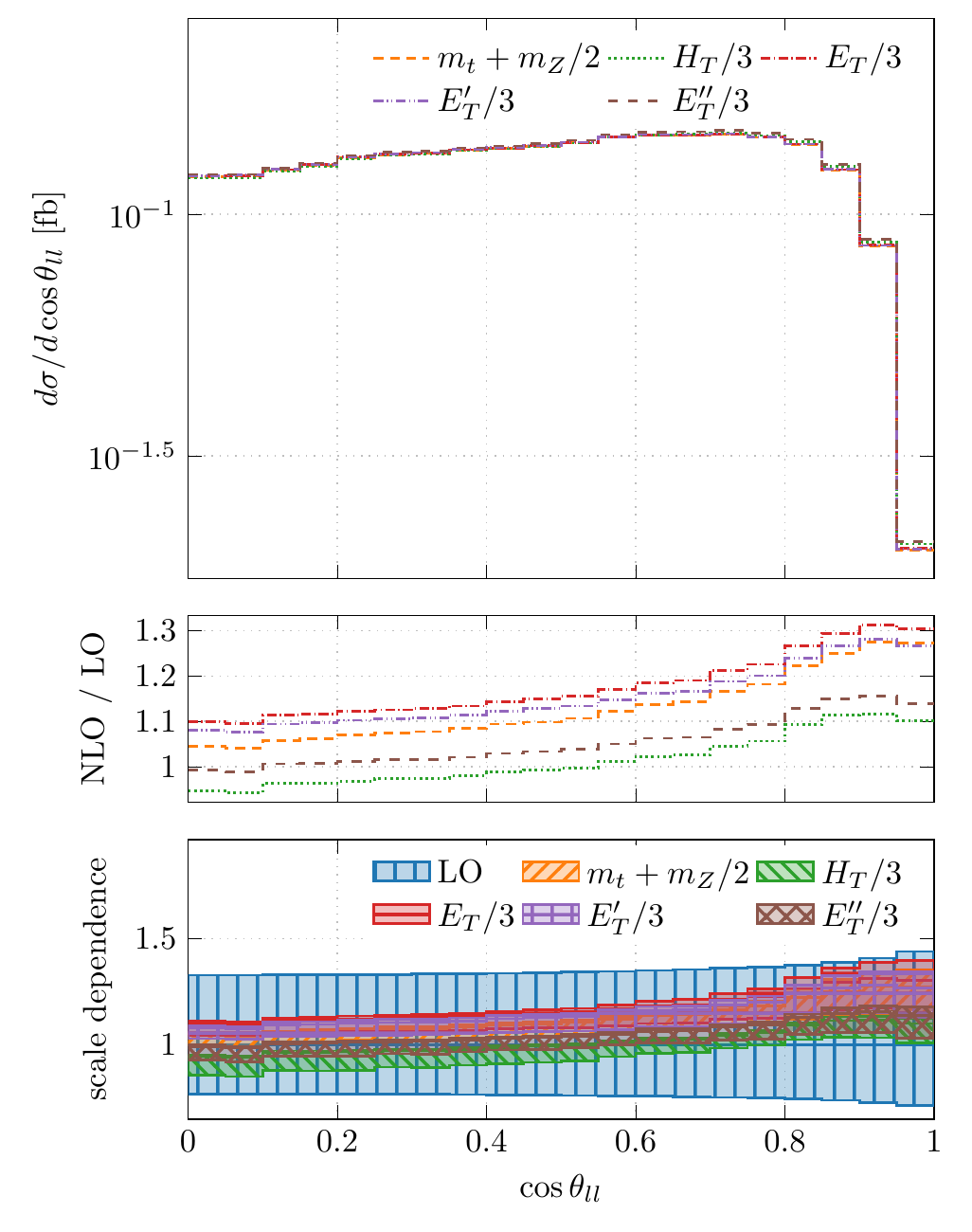} 
\caption{ Differential cross sections for $pp \to b \bar{b} e^+ \nu_e \mu^- \bar{\nu}_\mu \nu_\tau \bar{\nu}_\tau \, + X$ ("$t\bar{t}Z(Z\to\nu_\ell\bar{\nu}_\ell)$") at $\sqrt{s} = 13$ TeV as a function of $p_{T,l}$ and $\cos\theta_{ll}$ (defined in the text) \cite{Bevilacqua:2019cvp}. \textit{Upper panels:} absolute NLO QCD predictions for different scale choices. \textit{Middle panels:} differential $K$-factors. \textit{Lower panels}: scale uncertainty bands. The LO band refers to the scale choice $\mu_R = \mu_F = m_t + m_Z/2$. Results are based on the CT14 PDF set \cite{Dulat:2015mca}. }
\label{fig:ttZ}
\end{center}
\end{figure}

We begin our discussion with the $t\bar{t}Z(Z \to \nu_\ell\bar{\nu}_\ell)$ process.
An important task, when dealing with higher-order corrections, is to identify an \textit{optimum} for the renormalization and factorization scales. It is well known that the choice of a well motivated dynamical scale can sensibly improve the perturbative stability of higher-order predictions. With this motivation at hand we have considered five different functional forms for the scales:
\begin{eqnarray}
\mu_1  & = & m_t + m_Z/2  \label{Eq:fixed_scale} \,, \\
\mu_2  & = & E_T/3 =  \left( m_{T,t} + m_{T,\bar{t}} + p_{T,Z} \right)/3 \label{Eq:dyn_scale_1} \,, \\ 
\mu_3  & = & E_T^{\prime}/3  =  \left( m_{T,t} + m_{T,\bar{t}} + m_{T,Z}  \right)/3 \label{Eq:dyn_scale_2} \,, \\ 
\mu_4  & = & E_T^{\prime\prime}/3  =  \left( m_{T,t} + m_{T,\bar{t}}  \right)/3  \label{Eq:dyn_scale_3} \,, \\
\mu_5  & = & H_T/3 =  \left( p_{T,e^+} + p_{T,\mu^-} + p_{T,b} + p_{T,\bar{b}} + p_{T}^{miss} \right)/3 \,,\label{Eq:dyn_scale_4} 
\end{eqnarray}
where $m_{T,i} = \sqrt{p_{T,i} + m_i^2}$ is the transverse mass, and intermediate resonances are reconstructed from their decay products using Monte Carlo truth. Eq.(\ref{Eq:fixed_scale}) is a fixed-scale choice commonly adopted in the literature. Eq.(\ref{Eq:dyn_scale_1}-\ref{Eq:dyn_scale_3}) are examples of "resonance-aware" scales, motivated by the expectation that top-quark and $Z$ resonances describe the bulk of the cross section, while Eq.(\ref{Eq:dyn_scale_4}) is an example of "agnostic" scale which makes no assumption on intermediate resonant states. Figure \ref{fig:ttZ} shows NLO differential cross sections as a function of the averaged transverse momentum of the charged leptons ($p_{T,l}$) and of $\cos\theta_{ll} = \tanh(\Delta\eta_{\ell\ell})$. The latter is of special interest in DM searches through simplified models, as its shape is sensitive to the $CP$-nature and mass of the DM mediator \cite{Haisch:2016gry}. The better performance of dynamical scales is evident in the $p_{T,l}$ distribution, where the fixed-scale result does not fit the LO theory uncertainties in the tail. Looking at the $\cos\theta_{ll}$ distribution, the scales $H_T/3$ and $E_T^{\prime\prime}/3$ are in slightly better shape although the differential $K$-factor is far from constant over the whole plotted range. 
In consequence, a simple rescaling of LO predictions by a global $K$-factor cannot guarantee an accurate description of shapes. This motivates the need for a full NLO calculation.

\begin{figure}[h!]
\begin{center}
\includegraphics[width=.42\textwidth, height=6.7cm]{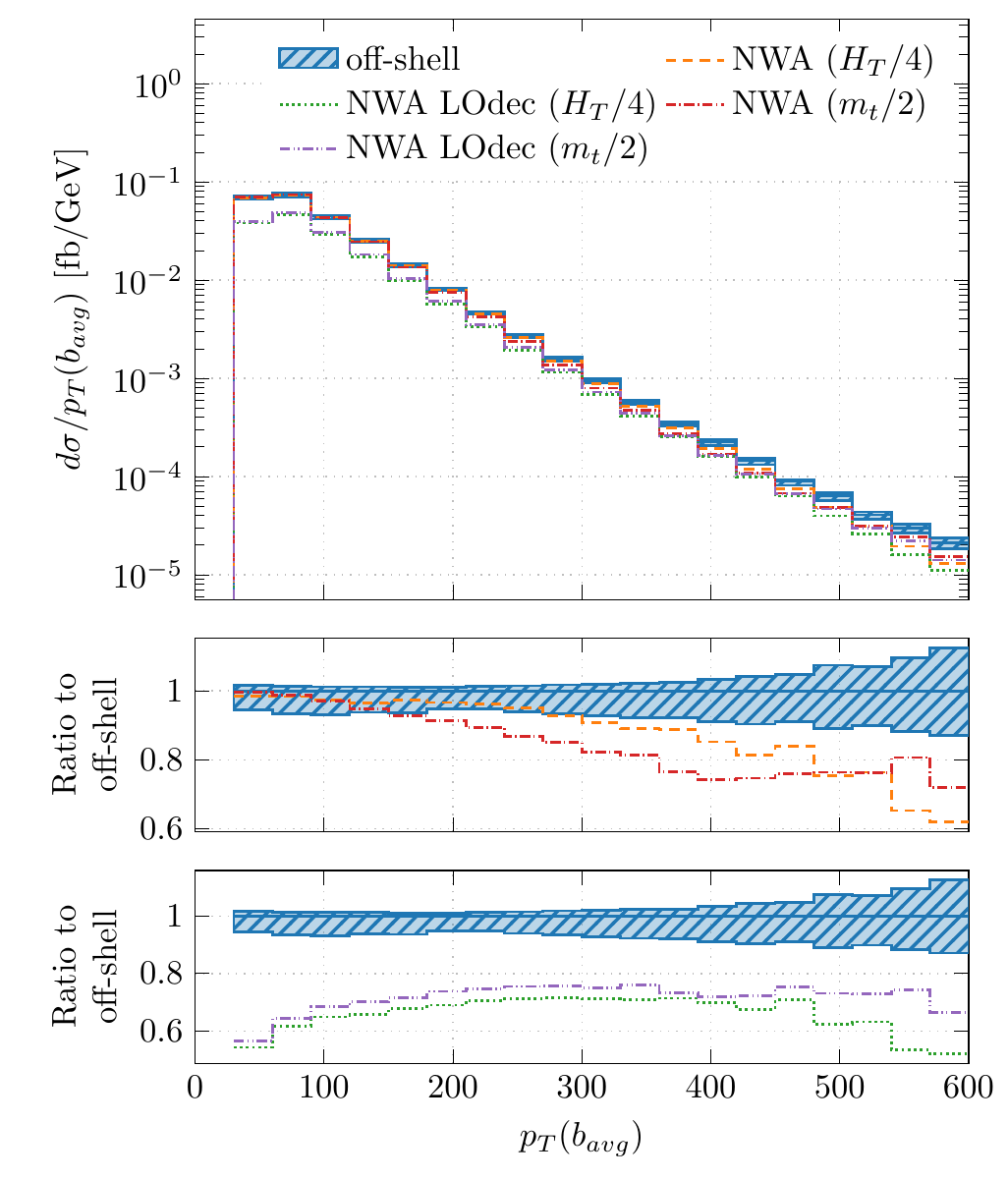} 
\hspace{0.3cm}
\includegraphics[width=.46\textwidth, height=6.7cm]{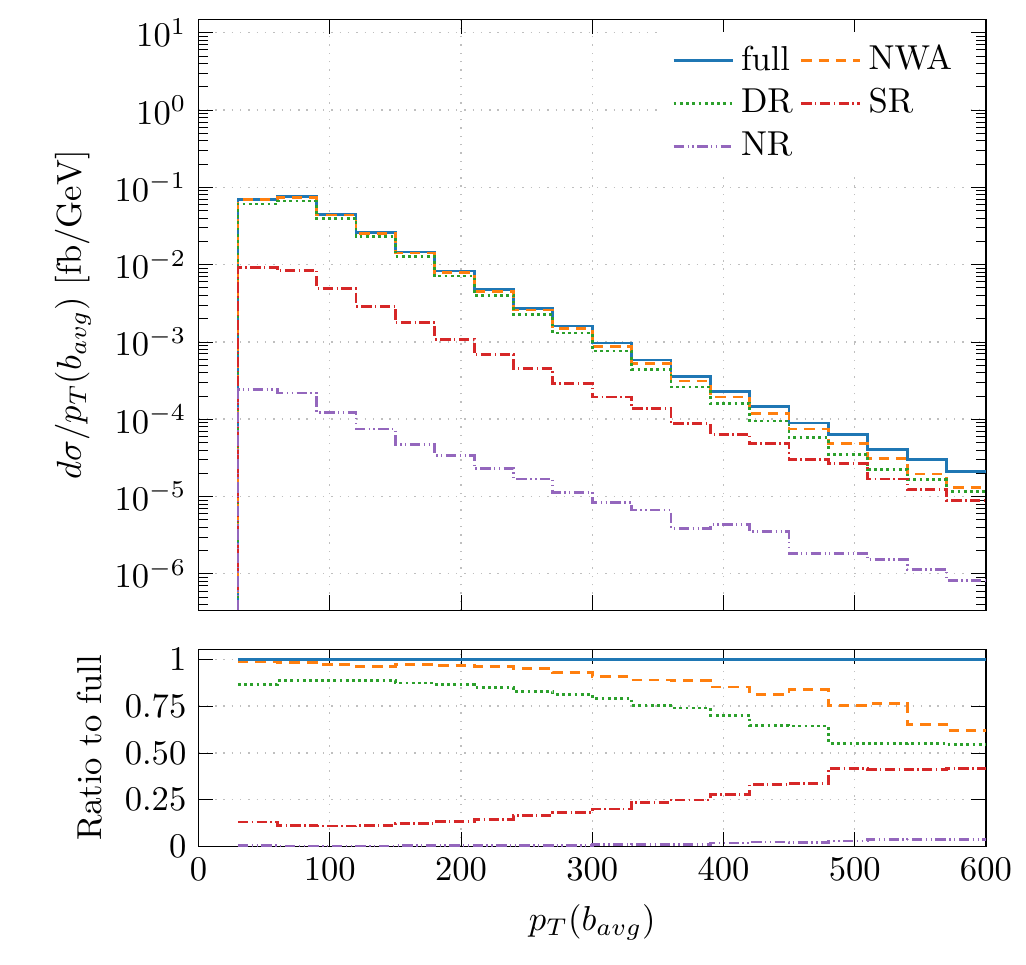}
\caption{Differential cross sections for $pp \to b \bar{b} e^+ \nu_e \mu^- \bar{\nu}_\mu \gamma \, + X$ ("$t\bar{t}\gamma$")  at $\sqrt{s} = 13$ TeV as a function of the average $p_T$ of $b$-jets \cite{Bevilacqua:2019quz}. The off-shell prediction is based on the scale choice $\mu_R = \mu_F = H_T/4$. All results are based on CT14 PDFs. The uncertainty bands refer to the off-shell calculation with default scale $H_T/4$. \textit{Left plot}: impact of different approaches for the modeling of top quark decays. \textit{Right plot}: impact of Double-, Single- and Non-Resonant contributions to the full NLO QCD calculation.}
\label{fig:ttA}
\end{center}
\end{figure}

In the next step we consider the $t\bar{t}\gamma$ process. We want to illustrate the impact of different approaches of  modeling top-quark decays. To this end we compare the full off-shell results against NWA with two different levels of accuracy, namely: (a) decays at NLO and photon radiation in both production and decays ("full NWA"); (b) decays at LO and photon radiation restricted to the production stage only ("$\mbox{NWA}_{\scriptsize \mbox{NLOdecay}}$"). The latter should mimic the expectation of a NLO+PS computation where the (spin-correlated) top decays do not include radiative effects. In Figure \ref{fig:ttA}, the NLO predictions for the average $p_T$ of the final state $b$-jets are reported. The uncertainty bands are estimated from the off-shell calculation via 7-point variation. We observe that $\mbox{NWA}_{\scriptsize \mbox{NLOdecay}}$ does not describe adequately the process in the whole  range. As discussed in \cite{Bevilacqua:2019quz}, also in agreement with earlier findings \cite{Melnikov:2011ta}, this is related to the fact that photon radiation from production and from decays provide comparable contributions to the total NLO cross section. Full NWA and off-shell results are in fair agreement up to $p_T(b_{avg})\sim 150$ GeV, while the discrepancy increases in the tail reaching $30-40\%$ depending on the scale considered. 
The increasing importance of the off-shell effects at high $p_T$'s can be understood by analysing the relative importance
of double-, single- and non-resonant contributions (denoted DR, SR and NR for brevity) to the full NLO calculation. 
These are extracted following a generalization of the method of Ref.\cite{Kauer:2001sp} which can be briefly sketched as follows: for each event, (i) we identify the most likely set of decay products for top and anti-top and reconstruct their  invariant masses, then (ii) we check how much the latters differ from the nominal top mass, $m_t$. If the difference lies within a predefined window, the (anti-)top quark is labeled "resonant", otherwise "non-resonant". Further details can be found in Ref.\cite{Bevilacqua:2019quz}.
The right plot in  Figure \ref{fig:ttA} shows a clear correlation between enhanced sensitivity to the off-shell effects and increasing importance of SR contributions. 

\section{Conclusions}

We have presented a full calculation of $pp \to t\bar{t}\gamma$ and $pp \to t\bar{t}Z(Z\to\nu_\ell\bar{\nu}_\ell)$ in the dilepton channel at $\sqrt{s} = 13$ TeV, including for the first time complete off-shell and non-resonant effects at NLO QCD accuracy. For the case of $t\bar{t}Z(Z\to\nu_\ell\bar{\nu}_\ell)$, we have shown that the presence of non-flat $K$-factors motivates the need of a NLO calculation for a proper description of shapes, which is crucial for specific DM analyses. For $t\bar{t}\gamma$ we have discussed a systematic comparison of various approaches of modeling top-quark decays, showing that radiative effects play an important role for a proper description of the process in the Narrow Width Approximation. Yet, the impact of off-shell contributions can reach several tens of percent in tails of several distributions of  interest.

%



\begin{thebibliography}{99}

\bibitem{Aaboud:2018hip}
  M.~Aaboud \textit{et al.} [ATLAS],
  \href{https://dx.doi.org/10.1140/epjc/s10052-019-6849-6}{\emph{Eur. Phys. J. C} \textbf{79} (2019) no.5, 382}
  [\href{https://arxiv.org/abs/1812.01697}{arXiv:1812.01697 [hep-ex]}].

\bibitem{Aad:2020axn}
  G.~Aad \textit{et al.} [ATLAS],
  \href{https://dx.doi.org/10.1007/JHEP09\%282020\%29049}{\emph{JHEP} \textbf{09} (2020), 049}
  [\href{https://arxiv.org/abs/2007.06946}{arXiv:2007.06946 [hep-ex]}].

\bibitem{Aaboud:2019njj}
  M.~Aaboud \textit{et al.} [ATLAS],
  \href{https://dx.doi.org/10.1103/PhysRevD.99.072009}{\emph{Phys. Rev. D} \textbf{99} (2019) no.7, 072009}
  [\href{https://arxiv.org/abs/1901.03584}{arXiv:1901.03584 [hep-ex]}].

\bibitem{Sirunyan:2017uzs}
  A.~M.~Sirunyan \textit{et al.} [CMS],
  \href{https://dx.doi.org/10.1007/JHEP08\%282018\%29011}{\emph{JHEP} \textbf{08} (2018), 011}
  [\href{https://arxiv.org/abs/1711.02547}{arXiv:1711.02547 [hep-ex]}].

\bibitem{CMS:2019too}
  A.~M.~Sirunyan \textit{et al.} [CMS],
  \href{https://dx.doi.org/10.1007/JHEP03\%282020\%29056}{\emph{JHEP} \textbf{03} (2020), 056}
  [\href{https://arxiv.org/abs/1907.11270}{arXiv:1907.11270 [hep-ex]}].

\bibitem{ATLAS:2020cxf}
 ATLAS Collaboration,
  \href{http://cdsweb.cern.ch/record/2725734}{ATLAS-CONF-2020-028}.

\bibitem{Baur:2004uw}
  U.~Baur, A.~Juste, L.~H.~Orr and D.~Rainwater,
  \href{https://dx.doi.org/10.1103/PhysRevD.71.054013}{\emph{Phys.\ Rev.\ D} {\bf 71} (2005) 054013}
  [\href{https://arxiv.org/abs/hep-ph/0412021}{hep-ph/0412021}].
  
\bibitem{Bylund:2016phk}
  O.~Bessidskaia Bylund, F.~Maltoni, I.~Tsinikos, E.~Vryonidou and C.~Zhang,
  \href{https://link.springer.com/article/10.1007/JHEP05(2016)052}{\emph{JHEP} {\bf 1605} (2016) 052}
  [\href{https://arxiv.org/abs/1601.08193}{arXiv:1601.08193 [hep-ph]}].

\bibitem{Schulze:2016qas}
  M.~Schulze and Y.~Soreq,
  \href{https://dx.doi.org/10.1140/epjc/s10052-016-4263-x}{\emph{Eur.\ Phys.\ J.\ C} {\bf 76} (2016) no.8, 466}
  [\href{https://arxiv.org/abs/1603.08911}{arXiv:1603.08911 [hep-ph]}].

\bibitem{Brivio:2019ius}
  I.~Brivio, S.~Bruggisser, F.~Maltoni, R.~Moutafis, T.~Plehn, E.~Vryonidou, S.~Westhoff and C.~Zhang,
  \href{https://dx.doi.org/10.1007/JHEP02\%282020\%29131}{\emph{JHEP} \textbf{02} (2020), 131}
  [\href{https://arxiv.org/abs/1910.03606}{arXiv:1910.03606 [hep-ph]}].

\bibitem{Bissmann:2019gfc}
  S.~Bissmann, J.~Erdmann, C.~Grunwald, G.~Hiller and K.~Kr\"oninger,
  \href{https://dx.doi.org/10.1140/epjc/s10052-020-7680-9}{\emph{Eur. Phys. J. C} \textbf{80} (2020) no.2, 136}
  [\href{https://arxiv.org/abs/1909.13632}{arXiv:1909.13632 [hep-ph]}].

\bibitem{PengFei:2009ph}
  P.~F.~Duan \textit{et al.},
  \href{https://journals.aps.org/prd/abstract/10.1103/PhysRevD.80.014022}{\emph{Phys.\ Rev.\ D} {\bf 80} (2009) 014022}
  [\href{https://arxiv.org/abs/0907.1324}{arXiv:0907.1324 [hep-ph]}].

\bibitem{PengFei:2011qg}
  P.~F.~Duan \textit{et al.},
  \href{https://dx.doi.org/10.1088/0256-307X/28/11/111401}{\emph{Chin.\ Phys.\ Lett.\ }  {\bf 28} (2011) 111401}
  [\href{https://arxiv.org/abs/1110.2315}{arXiv:1110.2315 [hep-ph]}].

\bibitem{Duan:2016qlc}
  P.~F.~Duan \textit{et al.},
  \href{https://www.sciencedirect.com/science/article/pii/S0370269317300035?via\%3Dihub}{\emph{Phys.\ Lett.\ B} {\bf 766} (2017) 102}
  [\href{https://arxiv.org/abs/1612.00248}{arXiv:1612.00248 [hep-ph]}].

\bibitem{Maltoni:2015ena}
  F.~Maltoni, D.~Pagani and I.~Tsinikos,
  \href{https://dx.doi.org/10.1007/JHEP02\%282016\%29113}{\emph{JHEP} {\bf 1602} (2016) 113}
  [\href{https://arxiv.org/abs/1507.05640}{arXiv:1507.05640 [hep-ph]}].

\bibitem{Melnikov:2011ta}
  K.~Melnikov, M.~Schulze and A.~Scharf,
  \href{https://dx.doi.org/10.1103/PhysRevD.83.074013}{\emph{Phys.\ Rev.\ D} {\bf 83} (2011) 074013}
  [\href{https://arxiv.org/abs/1102.1967}{arXiv:1102.1967 [hep-ph]}].
  
 \bibitem{Kardos:2014zba}
  A.~Kardos and Z.~Trocsanyi,
  \href{https://link.springer.com/article/10.1007\%2FJHEP05\%282015\%29090}{\emph{JHEP} {\bf 1505} (2015) 090}
  [\href{https://arxiv.org/abs/1406.2324}{arXiv:1406.2324 [hep-ph]}].

\bibitem{Lazopoulos:2008de}
  A.~Lazopoulos \textit{et al.}, 
  \href{https://www.sciencedirect.com/science/article/pii/S0370269308007843?via\%3Dihub}{\emph{Phys.\ Lett.\ B} {\bf 666} (2008) 62}
  [\href{https://arxiv.org/abs/0804.2220}{arXiv:0804.2220 [hep-ph]}].

\bibitem{Rontsch:2014cca}
  R.~R\"ontsch and M.~Schulze,
  \href{https://dx.doi.org/10.1007/JHEP07\%282014\%29091}{\emph{JHEP} {\bf 1407} (2014) 091,  Erratum: [\emph{JHEP} {\bf 1509} (2015) 132]}
  [\href{https://arxiv.org/abs/1404.1005}{arXiv:1404.1005 [hep-ph]}].

\bibitem{Garzelli:2011is}
  M.~V.~Garzelli, A.~Kardos, C.~G.~Papadopoulos and Z.~Trocsanyi,
  \href{https://journals.aps.org/prd/abstract/10.1103/PhysRevD.85.074022}{\emph{Phys.\ Rev.\ D} {\bf 85} (2012) 074022}
  [\href{https://arxiv.org/abs/1111.1444}{arXiv:1111.1444 [hep-ph]}].

\bibitem{Frixione:2015zaa}
  S.~Frixione \textit{et al.},
  \href{https://link.springer.com/article/10.1007/JHEP06(2015)184}{\emph{JHEP} {\bf 1506} (2015) 184}
  [\href{https://arxiv.org/abs/1504.03446}{arXiv:1504.03446 [hep-ph]}].

\bibitem{Kulesza:2020nfh}
  A.~Kulesza, L.~Motyka, D.~Schwartl\"ander, T.~Stebel and V.~Theeuwes,
  \href{https://dx.doi.org/10.1140/epjc/s10052-020-7987-6}{\emph{Eur. Phys. J. C} \textbf{80} (2020) no.5, 428}
  [\href{https://arxiv.org/abs/2001.03031}{arXiv:2001.03031 [hep-ph]}].

\bibitem{Broggio:2019ewu}
  A.~Broggio, A.~Ferroglia, R.~Frederix, D.~Pagani, B.~D.~Pecjak and I.~Tsinikos,
  \href{https://dx.doi.org/10.1007/JHEP08\%282019\%29039}{\emph{JHEP} \textbf{08} (2019), 039}
  [\href{https://arxiv.org/abs/1907.04343}{arXiv:1907.04343 [hep-ph]}].

\bibitem{Bevilacqua:2018woc}
  G.~Bevilacqua, H.~B.~Hartanto, M.~Kraus, T.~Weber and M.~Worek,
  \href{https://dx.doi.org/10.1007/JHEP10\%282018\%29158}{\emph{JHEP} {\bf 1810} (2018) 158}
  [\href{https://arxiv.org/abs/1803.09916}{arXiv:1803.09916 [hep-ph]}].

\bibitem{Bevilacqua:2018dny}
  G.~Bevilacqua, H.~B.~Hartanto, M.~Kraus, T.~Weber and M.~Worek,
  \href{https://dx.doi.org/10.1007/JHEP01\%282019\%29188}{\emph{JHEP} {\bf 1901} (2019) 188}
  [\href{https://arxiv.org/abs/1809.08562}{arXiv:1809.08562 [hep-ph]}].

\bibitem{Bevilacqua:2019cvp}
  G.~Bevilacqua, H.~B.~Hartanto, M.~Kraus, T.~Weber and M.~Worek,
  \href{https://dx.doi.org/10.1007/JHEP11\%282019\%29001}{\emph{JHEP} \textbf{11} (2019), 001}
  [\href{https://arxiv.org/abs/1907.09359}{arXiv:1907.09359 [hep-ph]}].

\bibitem{Bevilacqua:2019quz}
  G.~Bevilacqua, H.~B.~Hartanto, M.~Kraus, T.~Weber and M.~Worek,
  \href{https://dx.doi.org/10.1007/JHEP03\%282020\%29154}{\emph{JHEP} \textbf{03} (2020), 154}
  [\href{https://arxiv.org/abs/1912.09999}{arXiv:1912.09999 [hep-ph]}].
  
\bibitem{Bevilacqua:2011xh}
  G.~Bevilacqua, M.~Czakon, M.~V.~Garzelli, A.~van Hameren, A.~Kardos, C.~G.~Papadopoulos, R.~Pittau and M.~Worek,
  \href{https://www.sciencedirect.com/science/article/pii/S0010465512003761?via\%3Dihub}{\emph{Comput.\ Phys.\ Commun.\ }  {\bf 184} (2013) 986}
  [\href{https://arxiv.org/abs/1110.1499}{arXiv:1110.1499 [hep-ph]}].

\bibitem{vanHameren:2009dr}
  A.~van Hameren, C.~G.~Papadopoulos and R.~Pittau,
  \href{https://dx.doi.org/10.1088/1126-6708/2009/09/106}{\emph{JHEP} {\bf 0909} (2009) 106}
  [\href{https://arxiv.org/abs/0903.4665}{arXiv:0903.4665 [hep-ph]}].

\bibitem{Czakon:2009ss}
  M.~Czakon, C.~G.~Papadopoulos and M.~Worek,
  \href{https://dx.doi.org/10.1088/1126-6708/2009/08/085}{\emph{JHEP} {\bf 0908} (2009) 085}
  [\href{https://arxiv.org/abs/0905.0883}{arXiv:0905.0883 [hep-ph]}].
  
\bibitem{Bevilacqua:2013iha}
  G.~Bevilacqua, M.~Czakon, M.~Kubocz and M.~Worek,
  \href{https://dx.doi.org/10.1007/JHEP10\%282013\%29204}{\emph{JHEP} {\bf 1310} (2013) 204}
  [\href{https://arxiv.org/abs/1308.5605}{arXiv:1308.5605 [hep-ph]}].

\bibitem{Alwall:2006yp}
  J.~Alwall {\it et al.},
  \href{https://www.sciencedirect.com/science/article/pii/S0010465506004164?via\%3Dihub}{\emph{Comput.\ Phys.\ Commun.\ }  {\bf 176} (2007) 300}
  [\href{https://arxiv.org/abs/hep-ph/0609017}{hep-ph/0609017}].

\bibitem{Antcheva:2009zz}
  I.~Antcheva {\it et al.},
  \href{https://www.sciencedirect.com/science/article/pii/S0010465509002550?via\%3Dihub}{\emph{Comput.\ Phys.\ Commun.\ }  {\bf 180} (2009) 2499}
  [\href{https://arxiv.org/abs/1508.07749}{arXiv:1508.07749 [physics.data-an]}].

\bibitem{Bern:2013zja}
  Z.~Bern \textit{et al.},
  \href{https://www.sciencedirect.com/science/article/pii/S0010465514000241?via\%3Dihub}{\emph{Comput.\ Phys.\ Commun.\ }  {\bf 185} (2014) 1443}
  [\href{https://arxiv.org/abs/1310.7439}{arXiv:1310.7439 [hep-ph]}].

\bibitem{Haisch:2016gry}
  U.~Haisch, P.~Pani and G.~Polesello,
  \href{https://dx.doi.org/10.1007/JHEP02\%282017\%29131}{\emph{JHEP} {\bf 1702} (2017) 131}
  [\href{https://arxiv.org/abs/1611.09841}{arXiv:1611.09841 [hep-ph]}] .

\bibitem{Kauer:2001sp}
  N.~Kauer and D.~Zeppenfeld,
  \href{https://dx.doi.org/10.1103/PhysRevD.65.014021}{\emph{Phys. Rev. D} \textbf{65} (2002), 014021}
  [\href{https://arxiv.org/abs/hep-ph/0107181}{arXiv:hep-ph/0107181 [hep-ph]}].

\bibitem{Dulat:2015mca}
  S.~Dulat, T.~J.~Hou, J.~Gao, M.~Guzzi, J.~Huston, P.~Nadolsky, J.~Pumplin, C.~Schmidt, D.~Stump and C.~P.~Yuan,
  \href{https://dx.doi.org/10.1103/PhysRevD.93.033006}{\emph{Phys. Rev. D} \textbf{93} (2016) no.3, 033006}
  [\href{https://arxiv.org/abs/1506.07443}{arXiv:1506.07443 [hep-ph]}].

\end{thebibliography}
\end{document}